\documentstyle[12pt,aaspp4,psfig]{article}
\slugcomment{Accepted by {\it The Astronomical Journal}}

\lefthead{Harlaftis et al.}
\righthead{Nova Ophiuchi 1997}

\begin{document}

\title{A Doppler Map and Mass-Ratio Constraint \\
       for the Black-Hole X-Ray Nova Ophiuchi 1977 \altaffilmark{1}}

\vspace{1cm}

\author{Emilios T. Harlaftis, Danny Steeghs, Keith  Horne}
\affil{School of Physics and Astronomy, University of St. Andrews, KY16 
9SS, Scotland, UK}
\affil{Electronic mail: (ehh, ds10, kdh1)@st-andrews.ac.uk}

\vspace{0.5cm}

\and

\author{Alexei V. Filippenko}
\affil{Department of Astronomy, University of California,
    Berkeley, CA 94720-3411}
\affil{Electronic mail: alex@astro.berkeley.edu}

\vspace{1cm}


\altaffiltext{1}{Based on observations obtained at the W. M. Keck 
Observatory.} 


\begin{abstract} 
We have reanalyzed Keck observations of Nova Oph 1977 to extend the
work done by Filippenko et al. (1997), who recently determined a mass
function $f(M_{\rm x}) = 4.86 \pm 0.13~M_{\odot}$ for the compact
object.  We constrain the rotational broadening, $\upsilon \sin i
\leq 79$ km s$^{-1}$, at the 90\% confidence level, which
gives a mass ratio $q \leq 0.053$.  The K-type companion star of Nova
Oph 1977 contributes 28--37\% of the light at red wavelengths.  The
abnormal \ion{Li}{1} $\lambda$6708 absorption line from the companion
star is not detected (EW $< 0.12$~\AA), in contrast to four other X-ray
binaries.  An H$\alpha$ Doppler image of the system shows emission
from the companion star in addition to the accretion disk.
\end{abstract}




%

\section{INTRODUCTION}

Significant progress in demonstrating the existence of compact dark
stars more massive than neutron stars (i.e., black holes) has been
made over the last decade by studying X-ray novae in quiescence.  In
some cases, these systems consist of late-type (F5--M2) secondary
stars orbiting black holes ($\sim$4--15~$M_{\odot}$) with periods
ranging from 5.1 hours to 6.5 days (see, for example, van Paradijs \&
McClintock 1995).  After the initial outburst of the accretion disk
around the black hole, the system returns to quiescence, allowing the
photospheric lines of the faint companion (secondary) star to be
detected.  Radial-velocity studies of the companion star (e.g.,
Casares, Charles, \& Naylor 1992) can then provide the mass function
of the black hole,

\begin{equation} f(M_{\rm x}) = \frac{PK_{\rm c}^{3}}{2\pi G} = 
\frac{M_{\rm x}~\sin^{3}i}{(1+q)^{2}} , \end{equation}

\noindent 
where $K_{\rm c}$ is the semi-amplitude of the radial
velocity of the companion star (mass $M_{\rm c}$), $M_{\rm x}$ is the
mass of the black hole, and $q$ is the mass ratio $M_{\rm c}/M_{\rm
x}$.  Moreover, the value of $q$ can be obtained from the rotational
broadening of the companion star, $\upsilon \sin i$ (e.g., Marsh,
Robinson, \& Wood 1994), using the relation

\begin{equation} \frac{\upsilon \sin i}{K_{\rm c}} = 
0.462 \left[ (1+q)^{2} ~q \right] ^{1/3} .  \end{equation}

\noindent
This formula is derived by assuming that the spin period of the companion
star is locked to the binary period and by using Paczy\'{n}ski's (1971)
relation for $R_{\rm c}/a$ and $q \ll 1$, where $a$ is the semimajor
axis of the orbit. 

As summarized by Harlaftis \& Filippenko (1997), the advent of the
10-m Keck telescopes has enabled measurement of the mass function and
mass ratio for black-hole X-ray transients at magnitude
$R\approx21$--21.5.  The objects studied in detail thus far are GRO
J0422+32 (Filippenko et al. 1995b; Harlaftis et al. 1997) and
GS~2000+25 (Filippenko et al. 1995a; Harlaftis, Horne, \& Filippenko
1996, hereafter HHF).  Most recently, Filippenko et al. (1997,
hereafter FMLBV) derived a mass function $f(M_{\rm x}) = 4.86 \pm
0.13~ M_{\odot}$ for the compact object in Nova Ophiuchi 1977 by
measuring the radial-velocity semi-amplitude of the K-type companion
star ($K_{\rm c} = 447.6 \pm 3.9$ km s$^{-1}$), consistent with an
independent estimate by Remillard et al. [1996; $f(M_{\rm x}) = 4.0
\pm 0.8~ M_{\odot}$].

The reader should refer to FMLBV for an overview of Nova Oph 1977, a
description of the Keck observations with the Low Resolution Imaging
Spectrometer (LRIS; Oke et al. 1995), and the derivation of the mass
function of the compact object.  A total of 18 long-slit CCD spectra
were obtained on three nights (1996 May 12, and 1996 July 13, 14; see
Table 1 for a log of observations). Here, we perform an analysis
similar to the one we undertook for GS~2000+25 (HHF), and further
refined for GRO~J0422+32 (Harlaftis et al. 1997), in order to
determine the rotational broadening of the companion star to Nova Oph
1977 (which gives the system's mass ratio) and to investigate Doppler
maps of the H$\alpha$ emission-line distribution.

\section{RADIAL VELOCITIES REVISITED}

We re-extracted the 18 spectra using the PAMELA software package.  The
advantages of the package include optimal extraction, efficient
clipping of cosmic rays, and production of statistical uncertainties
based on photon counts and CCD readout noise (Horne 1986).  The
detailed reduction procedure is similar to that presented in HHF.

We dereddened the object spectra by $E$(\bv) = $0.9 $ mag. Although
Griffiths et al. (1978) derived $E$(\bv) $\approx 0.5$ mag during
outburst, use of their value still produced significantly red spectra,
yet we would expect a blue continuum because the accretion disk is the
dominant light source (see \S~3). Considering the small size of the
observed wavelength range, on the other hand, we cannot reliably
improve upon the Griffiths et al. reddening determination.  Since our
conclusions are independent of the precise value of the adopted
reddening, here we do not further consider it.

Table 2 gives nightly averages of the continuum flux density in the
range 6600--6800~\AA, the H$\alpha$ emission-line flux integrated
between --1900 and +1900 km s$^{-1}$ of line center, and the H$\alpha$
equivalent width (EW).  As already remarked by FMLBV, the spectrum
(continuum and H$\alpha$ profile) changed considerably between the May
and July epochs.  For example, a typical double-peaked H$\alpha$
profile (with a separation of $980 \pm 40$ km s$^{-1}$) was present in
May, but the line profile on July 14 was unusual, had a much smaller
EW, and included a narrow emission line at 6578~\AA\ ($\sim +700$ km
s$^{-1}$) whose origin is other than the sky.  This feature was also
observed at another epoch (July 1994) which suggests that it may be
related to the unusual shape of the profile (see Fig. 2 in Remillard
et al. 1996).  Interestingly, [N~II] $\lambda$6583 is +270 km s$^{-1}$
away, but the absence of evidence for [N~II] $\lambda$6548 makes this
an unlikely identification.

Results of the new radial-velocity analysis are collected in Table 3
and the salient points are as follows.  For K3/4, K5, K7, and M0
template stars, we find $438 < K_{\rm c} < 441$ km s$^{-1}$, and $-75
< \gamma < -31$ km s$^{-1}$. Note that the derived value of $K_{\rm
c}$ is independent of the template star used. The $\gamma$ velocities
span a larger range than expected from the uncertainties, perhaps
indicating that some of the heliocentric radial velocities of the
template stars (taken from the SIMBAD database) are erroneous. We
choose the average of the values found for the different templates and
use the spread to estimate its uncertainty: $\gamma = -54 \pm 15$ km
s$^{-1}$.  This is consistent with the result of FMLBV (average
$\gamma = -41.1 \pm 0.8$ km s$^{-1}$), but the uncertainty is now
considerably more realistic.

We adopt the K5 spectral type, which gives a radial-velocity ephemeris 
(see \S~4)

\begin{equation} 
V_R = \gamma + K_{\rm c} \ \left[ \sin \frac{2\pi}{P}(t-T_0) \right],
\end{equation}

\noindent 
with $K_{\rm c} = 441 \pm 6$ km s$^{-1}$, $\gamma = -54\pm
15$ km s$^{-1}$, $P=0.5228\pm0.0044$~d, and $T_0 = {\rm
HJD}~2,450,212.9800 \pm 0.0092$ (inferior conjunction of the companion
star).  For comparison, FMLBV found $K_{\rm c} = 447.6 \pm 3.9$ km
s$^{-1}$, $P = 0.5229 \pm 0.0044$~d, and $T_0 = {\rm
HJD}~2,450,212.9641 \pm 0.0038$ (after conversion to the phase
convention used here). They adopted 0.5229~d as the best of five periods
from a global $\chi^{2}$ analysis, and assigned an uncertainty of
0.0044~d, equal to the separation between adjacent periods.  We tried
the orbital solution with $P = 0.5272$~d to test for sensitivity of
our results, but found no significant difference in the $\chi^{2}$ or
$\upsilon \sin i$ compared to those obtained with $P = 0.5228$~d.

\section{ROTATIONAL BROADENING OF THE COMPANION STAR}

The spectral type, rotational broadening $\upsilon \sin i$, and
fractional contribution $f$ of the companion star to the total flux
can be determined by using the $\chi^{2}$ statistic after subtracting
different template spectra from the Doppler-corrected average spectrum
(see \S~5 in HHF, and references therein).  The steps of this analysis
were as follows.

The spectra were rebinned onto a common logarithmic wavelength scale
using a ``$\sin{x}/x$'' interpolation scheme to minimize data
smoothing (Stover et al. 1980).  This step provided spectra with a
uniform pixel size of 30.1 km s$^{-1}$, and was used to effectively
remove Doppler shifts arising from the Earth's motion, heliocentric
radial velocities of the template stars, and orbital motion of the
companion star as computed from the appropriate sinusoidal radial
velocity curve (see Table 3).

We then fitted low-order spline functions to the continua, and
normalized the spectra by dividing by the fitted continua and
subtracting 1.  In performing the continuum fits, all points above
$+3\sigma$ and below $-1.8\sigma$ were rejected to make the fits pass
close to the continuum rather than being biased downward by numerous
absorption lines and TiO bands.  This ``$\sigma$-clipping'' procedure
worked well on the template star spectra, but did not reject enough
absorption lines when fitting the individual spectra of Nova Oph 1977,
which had lower signal-to-noise (S/N) ratios. We therefore averaged
the Nova Oph 1977 spectra to increase the S/N ratio by a factor of
$\sim~4$.  The shape of the continuum obtained from a $\sigma$-clipped
spline for the mean spectrum was then scaled to fit (and subsequently
normalize) the individual spectra of Nova Oph 1977.

We then averaged the Doppler-corrected spectra obtained on 14 July,
which had the best S/N ratio and covered half an orbital cycle.  In
this mean spectrum, absorption features from the companion star are
relatively sharp because the Doppler shifts due to orbital motion have
been shifted to zero velocity.  The lines are still broadened,
however, by the rotation of the star, by changes in the orbital
velocity during the individual exposures, and by the instrumental
resolution of the spectrograph.  The instrumental profile (full width
at half-maximum [FWHM] = 108 km s$^{-1}$) is of course the same for
the template star and Nova Oph 1977 spectra.  Our procedure to
estimate the companion star's rotational broadening parameter
$\upsilon \sin i$ is therefore to (a) ``blur'' the template star
spectra, to simulate rotational broadening and the residual orbital
drifts, (b) scale each blurred template spectrum by a factor $f$, to
fit the mean spectrum of Nova Oph 1977, and finally (c) use the
$\chi^2$ statistic to locate the parameter values that give the best
fit.

The line broadening function we adopted is the convolution of the
rotational broadening profile of a limb-darkened spherical star with
rectangular profiles that simulate the velocity drifts during the
individual exposures.  The velocity drift during an exposure of
duration $\Delta t$ is

\begin{equation}
\Delta V \approx \left| \frac{\partial V_R}{\partial t} \right| \Delta t
	= \frac{ 2\pi K_{\rm c} \Delta t }{P} \left| \cos{\phi} \right|.
\end{equation}

\noindent
Table~1 lists the exposure durations $\Delta t~\approx$ 25--30 min,
the binary phases $\phi$, and the corresponding velocity drifts $\Delta
V$, which range from 5 to 92 km s$^{-1}$.  The template spectra were
convolved with rectangular profiles corresponding to the orbital
drifts during each exposure, and then averaged using weights identical
to those used to obtain the mean spectrum of Nova Oph 1977.
 
We adopted the rotational broadening profile of a spherical star (Gray
1976) with $\upsilon \sin i$ ranging from 0 to 150 km s$^{-1}$ in
steps of 10 km s $^{-1}$, and limb-darkening coefficient $u=0.5$.  Our
data have insufficient quality to support independent estimation of
$u$ and $\upsilon \sin i$.  We note that the systematic error
resulting from $u=0.5\pm0.5$ is $\pm 7\%$ in $\upsilon \sin i$,
smaller than the statistical uncertainties.  Marsh, Robinson, \& Wood
(1994) investigated the effect of the Roche-lobe distortion on the
rotational broadening.  Their results show that the error we incur by
adopting a spherical star is also less than the statistical noise in our
$\upsilon \sin i$ estimates.

We scaled each blurred template spectrum by a factor $0 < f < 1$ to
match the absorption-line strengths in the Doppler-corrected mean
spectrum of Nova Oph 1977, and subtracted one from the other to obtain
a spectrum of residuals.  At this stage the residuals spectrum showed
long-scale trends resulting from small differences in the continuum
fits, and short-scale residuals around the absorption lines resulting
from the slightly inaccurate fits of the orbital velocity amplitude
and rotational broadening parameters.  We eliminated the long-term
component by applying a high-pass Gaussian filter of FWHM = 900 km
s$^{-1}$ (30 pixels), and then computed $\chi^{2}$ from the high-pass
filtered residuals.

Next, values of $\upsilon \sin i$ and $f$ were chosen to minimize
$\chi^{2}$, with results as given in Table 4.  At the 90\% confidence
level, all templates place an upper limit on $\upsilon \sin i$ below
79 km s$^{-1}$.  At the 68\% confidence level, the templates give
upper limits with the exception of the K5 template, which gives a
best fit at $\upsilon \sin i = 50^{+17}_{-23}$ km s$^{-1}$.  This is
illustrated in Figure 1, which displays $\chi^{2}$ versus $\upsilon
\sin i$ for the K5 template.  The plot shows a non-Gaussian
distribution ($\chi^{2}$ becomes flatter as $\upsilon \sin i$
approaches 0, caused by the finite instrumental resolution) with the
above-mentioned minima at the 68\% and 90\% confidence levels (Lampton,
Margon, \& Bowyer 1976).  Table 4 also shows that the companion star
contributes $\sim$ 26--37\% of the light at red wavelengths; the low
end of this range is consistent with the high end of that given by
FMLBV (10--30\%). 

The $\chi^2$ fitting techniques can be used to estimate the spectral
type of the companion star, but the best-fit $\chi^2$ values are not
necessarily a simple function of spectral type because different noise
levels affect our template star spectra, We found that the global
$\chi^{2}$ (sum of 18 values of $\chi^{2}$, one for each spectrum of
Nova Oph 1977) is similar for all the K-type templates and slightly
better than for the M0 template.  However, in the average spectrum of
Nova Oph 1977 there is no trace of the TiO bands typical of M-type
stars. Since the K5~V template gives a clear minimum in the plot of
$\chi^2$ versus $\upsilon \sin i$ (Fig. 1), we adopt a spectral
type of K5 for the companion star. This is consistent with the
conclusion of FMLBV, who chose K7 but gave a range of K3--M0.

We also analyzed the data using a more refined technique, which we
applied previously to GS~2000+25 (HHF) and GRO~J0422+32 (Harlaftis et
al. 1997).  This technique is basically the same as that described
above, except that instead of fitting to a single Doppler-corrected
average spectrum of Nova Oph 1977, we subtracted our shifted, blurred, 
and scaled template star spectra from each of the individual spectra of
Nova Oph 1977, using the same $\upsilon \sin i$ but 18 different scale
factors $f$ for the 18 individual spectra.  The plot of global
$\chi^{2}$ versus $\upsilon \sin i$ is nearly identical to that
derived from analysis of the mean spectrum, and serves to confirm our
results.  For comparison, the $\upsilon \sin i$ derived for the K5
template is also $50^{+18}_{-26}$ km s$^{-1}$.  The spectrum at phase
0.25 (smearing of only 5 km s$^{-1}$ during the exposure) gives
$\upsilon \sin i = 70^{+31}_{-53}$ km s$^{-1}$.  The individual $f$
values are also consistent within $1\sigma$ with those in Table 4.

\section{THE K5~V SUBTRACTED SPECTRUM}

The procedure applied to the Doppler-corrected average spectrum of
Nova Oph 1977 using the K5~V template is illustrated in Figure 2.  The
spectrum of the K5~V template (HD~125354) is shown at the bottom,
binned to 124 km s$^{-1}$ pixels.  This template was processed as
described in \S~3, and a rotational broadening profile corresponding
to $\upsilon \sin i = 50$ km s$^{-1}$ was applied (see Table 4).  The
result, after multiplication by $f=0.37$ (Table 4), is the second
spectrum from the bottom in Figure 2. The spectrum above it is the
Doppler-corrected average of the Nova Oph 1977 data in the rest frame
of the companion star. Finally, the residual spectrum shown at the top
is the result of subtracting the simulated K5~V template from the
Doppler-shifted average spectrum. 

The K-star absorption lines are evident in the Doppler-corrected
average, and they are almost completely removed by subtraction of the
template spectrum (e.g., the Na~I~D and 6495 \AA \ lines).  The
residual spectrum brings out disk lines and possible anomalous
strengths of companion star lines not removed by the template.
Emission from \ion{He}{1} $\lambda$5876 and $\lambda$6678 is absent.
There is also no clear evidence for \ion{Li}{1} $\lambda$6708
absorption to a $1\sigma$ EW upper limit of 0.12~\AA\ relative to the
observed continuum; this is formally consistent with the marginal
detection suggested by FMLBV (EW $= 0.08 \pm 0.04$~\AA).

\section{THE H$\alpha$ DOPPLER MAP}

We used the maximum entropy Doppler tomography method (Marsh \& Horne
1988; Harlaftis \& Marsh 1996) to reconstruct a Doppler map of
H$\alpha$ emission from the set of spectra obtained on July 14
(Fig. 3), when half of an orbital cycle was covered.  The trailed
spectra in the top panel show a weak double-peaked profile with an
``S-wave" component moving from red to blue at phase 0.9.  The orbital
phases were determined from the ephemeris given in \S~2. The double
peak is a classical signature of an accretion disk. The constructed
tomogram is the simplest image that can reproduce (estimated by the
$\chi_{\nu}^{2}$ statistic) our observed line profiles at each
phase. The iterative process of building up the image is repeated
until convergence at the maximum entropy solution is reached at the
desired $\chi_{\nu}^{2}$ value.  The final Doppler image was then used
to build the predicted spectra by projecting the image at the observed
binary phases (see bottom panel in Fig. 3 for computed spectra).

The Doppler image shows the weak ring-like distribution of an
accretion disk (which projects to form the double-peaked line
profiles) with emission present at velocities ranging between 300 and
1000 km s$^{-1}$ (center panel of Fig. 3).  The narrow line at
6577~\AA\ causes a narrow ring in the image. At phase 0.9, the effect
of the strong ``S-wave" component in the spectrum is evident on the
image.  The path of Keplerian velocities along the gas stream (upper
path) and the ballistic trajectory (lower path) are plotted along with
the Roche lobe of the companion star (for $q = 0.034$, $K_{\rm x} =
16$ km s$^{-1}$).  The image suggests that there is emission related
to the companion star and possibly the gas stream.  The H$\alpha$
emission line from the companion star is also seen in Figure 2 (top
spectrum).  The linear intensity scale is common for all panels.

\section{DISCUSSION}

The velocity semi-amplitude $K_{\rm c} = 441 \pm 6$ km s$^{-1}$ and
period $P = 0.5228 \pm 0.0044$~d imply a mass function

\begin{equation} f(M_{\rm x}) =  (4.65 \pm 0.21)~ M_{\odot} = 
\frac{M_{\rm x}~\sin^{3}i}{(1+q)^{2}} , \end{equation}

\noindent consistent with the FMLBV estimate of $4.86 \pm
0.13~M_{\odot}$.  The constraint $\upsilon \sin i \leq 79$ km s$^{-1}$
(and with a likely value of $\upsilon \sin i = 50^{+17}_{-23}$ km
s$^{-1}$ for a K5 companion star) gives a mass ratio $q~\leq 0.053$
($q = 0.014^{+0.019}_{-0.012}$ for K5) using equation (2). The mass
ratio implies $K_{\rm x}=qK_{\rm c} \leq 23$ km s$^{-1}$ ($K_{\rm x} =
6^{+9}_{-5}$ km s$^{-1}$ for K5).  Note that FMLBV obtained $K_{\rm x}
= 4.5 \pm 1.8$ km s$^{-1}$ (and $q = 0.01$) by measuring the centroid
of the H$\alpha$ emission line.  This makes Nova Oph 1977 one of the
most extreme mass-ratio systems, comparable to GS~2000+25 ($q = 0.042
\pm 0.012$; HHF).  

Remillard et al. (1996) constrained the system
inclination to be 60$^{\circ} < i < 80^{\circ}$, based on the
ellipsoidal variations due to the companion star on the one hand and
the absence of eclipses on the other hand.  For $q = 0.014$ (the
formal mass ratio obtained from $\upsilon \sin i = 50$ km s$^{-1}$),
the resulting masses of the compact object and companion star are
respectively $5.0 < M_{\rm x} < 7.4 ~M_{\odot}$ and $0.07 < M_{\rm c}
< 0.10~M_{\odot}$ (with a contribution of 26--37\% of the light in our
red spectra).  For comparison, a normal K5~V star has a mass of
$0.39~M_{\odot}$ (Allen 1976). An upper limit to the companion star
mass from our upper limit to the mass ratio ($q~\leq 0.053$) gives
$4.9 < M_{\rm x} < 7.9 ~M_{\odot}$ and $0.26 < M_{\rm c} <
0.42~M_{\odot}$ for $q$=0.053.  Undermassive (i.e., evolved) companion
stars are not uncommon in black-hole candidates (e.g., Harlaftis et
al. 1997, and references therein). McClintock \& Remillard (1990) also
infer a very low mass companion ($\sim 0.1 ~M_{\odot}$) in the
neutron-star binary Cen X-4.

We do not detect the \ion{Li}{1} $\lambda$6708 line (EW $< 0.12$~\AA;
$1\sigma$) which has been observed in four X-ray binaries that have
similar system properties (e.g., EW= 0.27 $\pm$ 0.04~\AA\ in
GS~2000+25; Filippenko et al. 1995a).  The absence of \ion{Li}{1} in
Nova Oph 1977 suggests that it may not be present in all X-ray
transients. A similar result was found for GRO~J0422+32 which,
however, has an M2 companion star (Filippenko et al. 1995b; Harlaftis
et al.  1997). In this case, the outburst mechanism (spallation close
to the black hole; Mart\'\i n et al. 1994) may not be a viable model
for the presence of lithium on the companion star, and some other
mechanism may be responsible.  Recent work on cool stars indicates
that the \ion{Li}{1} abundance may be linked to stellar activity
rather than age (Soderblom et al. 1993). The high rotation rates of
the companion stars in X-ray transients and tidal effects caused by
the primary star may contribute more to Li enhancement.  Clearly, we
need confirmation of such a low abundance of \ion{Li}{1} in a
different K-star/black-hole X-ray binary.

Further progress on Nova Oph 1977 requires higher resolution spectra,
particularly at orbital phases 0.25 and 0.75 (where the smearing is
minimal) to improve the estimate of the rotational broadening and
hence the mass ratio.  The inclination may also be further constrained
by modeling the ellipsoidal modulations of the companion star at
infrared wavelengths.

\acknowledgments

The W. M. Keck Observatory, made possible by the W. M. Keck Foundation, is
operated as a scientific partnership between Caltech and 
the University of California.  We thank the observatory
staff, as well as T. Matheson, D. C. Leonard,
A. J. Barth, and S. D. Van Dyk, for their assistance.
The data analysis was carried out at the STARLINK network
(St. Andrews node, UK). Use of software developed mainly by T. Marsh is 
acknowledged. This research has
benefitted from the SIMBAD database, operated at CDS, Strasbourg, France.
AVF is grateful for support through NSF grant AST--9417213.

\clearpage
\begin{deluxetable}{rccccrrr}
\tableheadfrac{0}
\tablecaption{Journal of Observations \label{tbl-1}}
\tablewidth{450pt}
\tablehead{
\colhead{N} & 
\colhead{1996 Date} & 
\colhead{UT(mid)} &
\colhead{$\Delta t$ (s)} &  
\colhead{HJD\tablenotemark{a}} & 
\colhead{$\phi$\tablenotemark{b}} & 
\colhead{$V_R$\tablenotemark{c} } &
\colhead{$\Delta V$\tablenotemark{d}} 
}
\startdata
 1 &12 May &10:32:44& 1800& 215.94463 &  5.6711 &$-407\pm 30 $ & 53\nl
 2 &12 May &11:03:24& 1800& 215.96594 &  5.7119 &$-511\pm 15 $ & 26\nl
 3 &12 May &12:06:14& 1500& 216.00957 &  5.7954 &$-500\pm 36 $ & 26\nl
 4 &12 May &12:31:56& 1500& 216.02742 &  5.8295 &$-426\pm 22 $ & 44\nl
 5 &13 July &11:47:21& 1500& 277.99614 &124.3714 &$ 211\pm 15 $ & 64\nl
 6 &13 July &12:14:10& 1630& 278.01476 &124.4070 &$ 159\pm 21 $ & 83\nl
 7 &14 July &06:30:37& 1500& 278.77614 &125.8639 &$-425\pm 13 $ & 61\nl
 8 &14 July &06:56:41& 1500& 278.79424 &125.8981 &$-329\pm 28 $ & 73\nl
 9 &14 July &07:22:22& 1500& 278.81208 &125.9322 &$-271\pm 15 $ & 84\nl
10 &14 July &07:52:13& 1500& 278.83281 &125.9719 &$-165\pm 16 $ & 91\nl
11 &14 July &08:17:54& 1500& 278.85064 &126.0060 &$ -49\pm 11 $ & 92\nl
12 &14 July &08:43:44& 1500& 278.86858 &126.0403 &$  73\pm 22 $ & 89\nl
13 &14 July &09:10:50& 1500& 278.88740 &126.0763 &$ 128\pm 21 $ & 82\nl
14 &14 July &09:36:38& 1500& 278.90532 &126.1106 &$ 197\pm 15 $ & 70\nl
15 &14 July &10:17:32& 1500& 278.93371 &126.1649 &$ 276\pm 42 $ & 47\nl
16 &14 July &10:45:46& 1800& 278.95332 &126.2024 &$ 328\pm 30 $ & 27\nl
17 &14 July &11:18:26& 1800& 278.97599 &126.2458 &$ 351\pm 19 $ & 5 \nl
18 &14 July &11:51:45& 2100& 278.99914 &126.2901 &$ 378\pm 19 $ & 32\nl
\enddata
    
\tablenotetext{a}{HJD--2,450,000 at midpoint of exposure.}
\tablenotetext{b}{Binary phase $\phi$, using $P = 0.5228$~d and $T_0 =
2,450,212.9800$.}  
\tablenotetext{c}{Radial velocities (km s$^{-1}$)
measured by cross-correlation against the K5 template.}
\tablenotetext{d}{Orbital broadening (km s$^{-1}$) using $K_{\rm
c}=441$ km s$^{-1}$.}  

\end{deluxetable}
    
\clearpage
    
\begin{deluxetable}{lrcccc}
\tablecaption{Average Properties \label{tbl-1}}
\tablewidth{430pt}

\tablehead{
\colhead{1996 Date} & 
\colhead{Spectra} & 
\colhead{HJD\tablenotemark{a}} & 
\colhead{$F_{\nu}$\tablenotemark{b}} & 
\colhead{Flux(H$\alpha$)\tablenotemark{c}} & 
\colhead{EW(H$\alpha$)\tablenotemark{d}} }
\startdata

12 May & 4   &215.97788 & 0.0351 $\pm$ 0.0002 &11.5 $\pm$ 0.2  & 65 $\pm$ 1 \nl        
13 July & 2   &278.00286 & 0.0117 $\pm$ 0.0002\tablenotemark{e} 
&1.00 $\pm$ 0.07\tablenotemark{e} & 17 $\pm$ 1 \nl              
14 July & 12  &278.85411 & 0.0196 $\pm$ 0.0001 &1.90 $\pm$ 0.05 & 18 $\pm$ 0.5 \nl  
\enddata

\tablenotetext{a}{HJD--2,450,000 at midpoint of night.}
\tablenotetext{b}{Continuum flux density in the range 6600--6800~\AA~(mJy).}
\tablenotetext{c}{H$\alpha$ emission-line flux (10$^{-15}$ erg s$^{-1}$ 
cm$^{-2}$).}
\tablenotetext{d}{Equivalent width of H$\alpha$ emission (\AA).}
\tablenotetext{e}{Not photometric value.}

\end{deluxetable}

\begin{deluxetable}{llrcc}
\tablecaption{Results of Radial Velocity Analysis \label{tbl-2}}
\tablewidth{420pt}

\tablehead{
\colhead{Star} & \colhead{Spectral type} & \colhead{$\gamma$ (km s$^{-1}$)} 
& \colhead{$K_{\rm c}$ (km s$^{-1}$)} & \colhead{$\chi^{2}$~($\nu$=14)}}
\startdata
HD~109333           & K3/4~V& $ -47\pm7 $    & $438 \pm 7 $  & 23 \nl
BD~$-05^\circ$3763  & K5~V  & $ -75\pm5 $    & $441 \pm 6 $  & 24 \nl
HD~125354           & K7~V  & $ -62\pm6 $    & $438 \pm 7 $  & 20 \nl
BD~+42$^\circ$2296  & M0~V  & $ -31\pm5 $    & $439 \pm 7 $  & 24 \nl
\enddata

\end{deluxetable}

\begin{deluxetable}{llccc}
\tablecaption{Optimal Subtraction of Companion Star \label{tbl-3}}
\tablewidth{420pt}

\tablehead{
\colhead{Star} & 
\colhead{Spectral type} & 
\colhead{$\upsilon \sin i$\tablenotemark{a} } & 
\colhead{$f$\tablenotemark{b}} & 
\colhead{$\chi^{2}$($\nu$=992)}
}
\startdata
HD~109333            & K3/4~V & $\leq 64$  &$0.26 \pm 3$  & 670 \nl
BD~$-05^\circ$3763  & K5~V   & $\leq 79$  &$0.37 \pm 4$  & 676 \nl
HD~125354            & K7~V   & $\leq 60$  &$0.30 \pm 3$  & 700 \nl
BD~+42$^\circ$2296  & M0~V   & $\leq 73$  &$0.30 \pm 3$  & 673 \nl
\enddata

\tablenotetext{a}{90\% confidence level.}
\tablenotetext{b}{Fraction of the light that is contributed by the 
companion star at 6250~\AA.}

\end{deluxetable}


\clearpage

\begin{figure}
\centerline{\psfig{figure=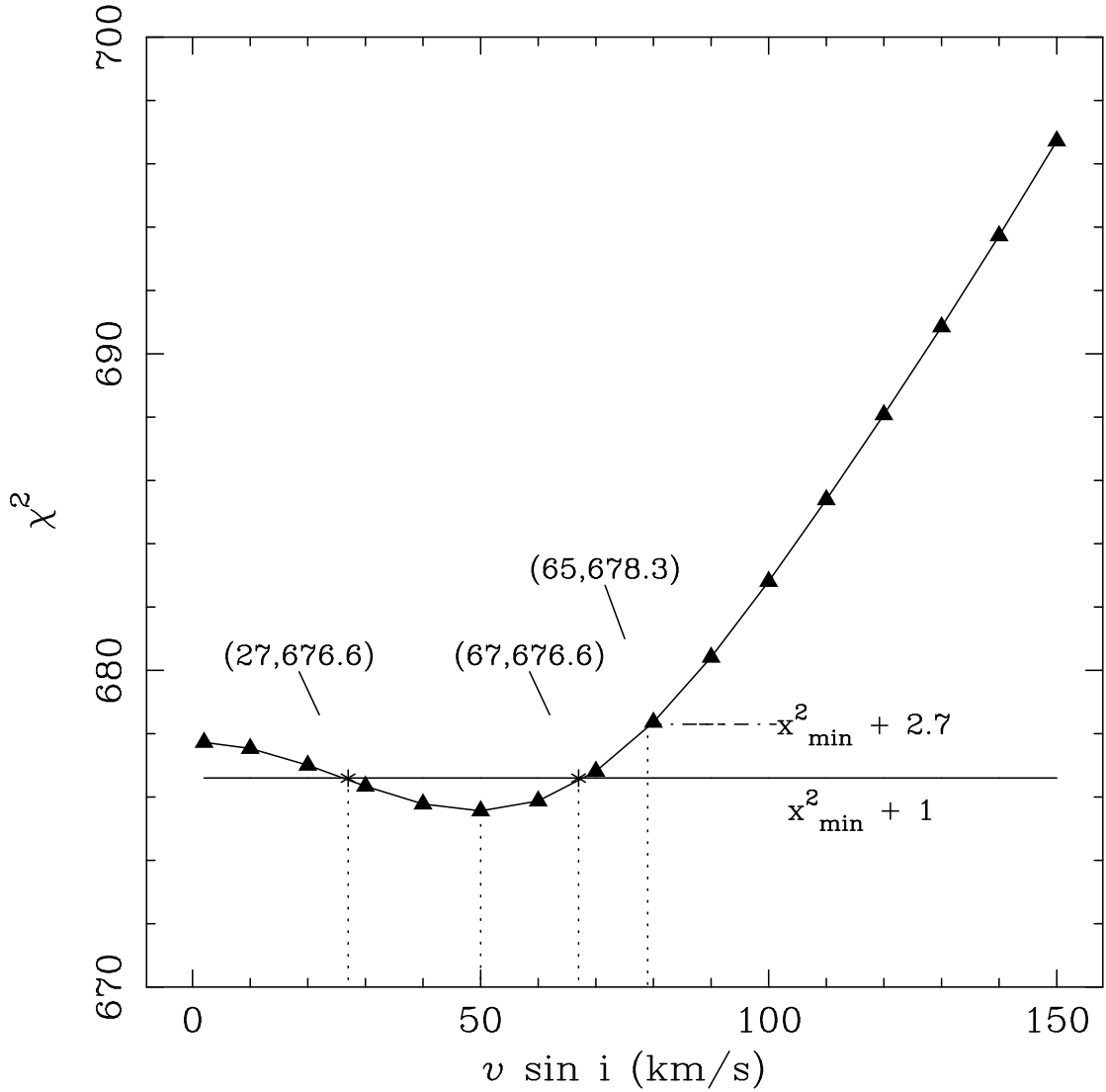,angle=-90,height=16cm}}
\caption[]{
The optimal subtraction of the main-sequence template from the Nova Oph 1977
spectra gives a $\chi^{2}$ distribution for a series of rotational broadenings
convolved with the template spectra. The $\chi^{2}$ distribution for
the K5 template (BD~$-05^{\circ}$3763) shows a minimum at 50 km s$^{-1}$.
The values for the 68\% ($\chi^{2}_{\rm min}$ + 1) and 90\% 
($\chi^{2}_{\rm min}$ + 2.7) confidence levels are marked.}
\end{figure}

\begin{figure}
\centerline{\psfig{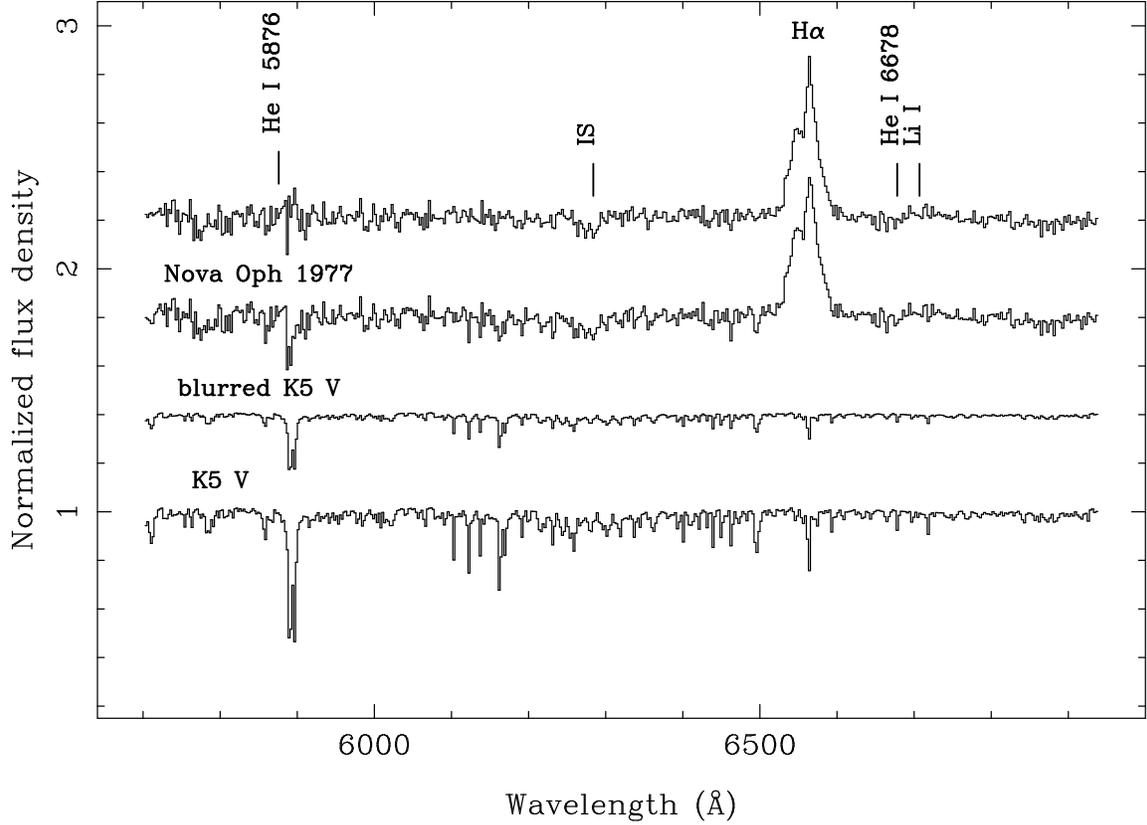}}
\caption[]{
Results of the technique followed to extract $\upsilon \sin i$ and $f$
for the companion star. From bottom to top are normalized spectra of
the K5~V template BD $-05^{\circ}3763$, the K5~V template convolved
with a complex profile to simulate effects of orbital smearing and
rotational broadening ($\upsilon \sin i = 50$ km s$^{-1}$), the
Doppler-corrected average spectrum of Nova Oph 1977, and the residual
spectrum after subtraction of the template star times $f = 0.37$. All
spectra are binned to 124 km s$^{-1}$ pixel$^{-1}$. An offset of 0.4
mJy was added to each successive spectrum for clarity. The residual
spectrum is essentially the disk spectrum (featureless continuum,
broad H$\alpha$ emission).  An interstellar line is marked (IS), as is
the expected position of the undetected \ion{Li}{1} $\lambda$6707.8
absorption.}
\end{figure}

\begin{figure}

\centerline{\psfig{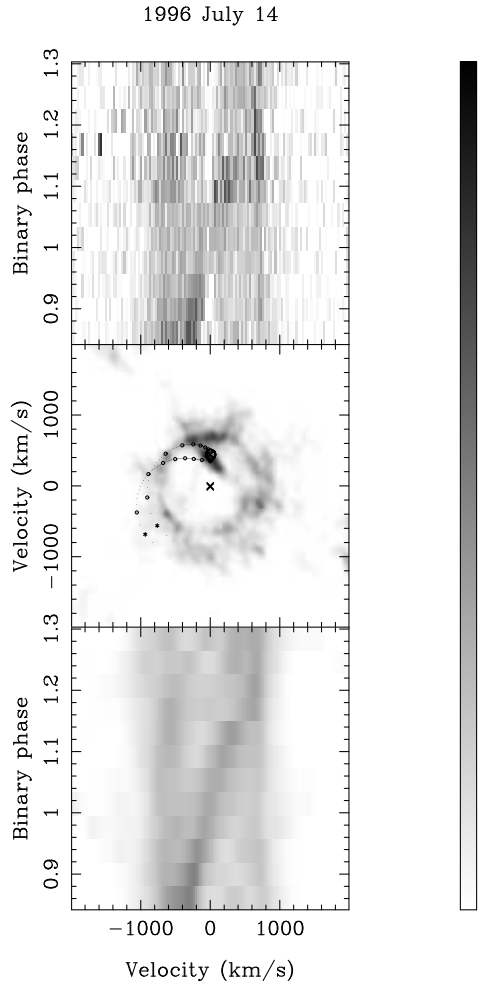}}
\caption[]{
Trailed spectra of H$\alpha$ on 1996 July 14 ({\it top panel}). 
The ordinate is orbital phase, while
the abscissa is velocity relative to the line center. A double-peaked
profile is discernible, as is an ``S-wave" component crossing from 
blue to red. The MEM Doppler map ({\it middle panel}) shows a
weak ring typical of accretion disk line distributions and identifies the 
companion star as the origin of the ``S-wave" component. 
Profiles projected from the image
are also displayed ({\it bottom panel}). See text for details.}
\end{figure}

\end{document}